\documentclass[12pt]{article}

\def\be{\begin{equation}}
\def\ee{\end{equation}}
\def\bea{\begin{eqnarray}}
\def\eea{\end{eqnarray}}

\title{\bf Are small hyperbolic universes observationally
detectable?}

\author{
G.I. Gomero$^{1,2}$\thanks{gomero@cbpf.br, chapita@gft.ucp.br}, \ 
M.J. Rebou\c{c}as$^1$\thanks{reboucas@cbpf.br}, \ 
R. Tavakol$^{1,3}$\thanks{r.tavakol@qmw.ac.uk} \\ 
\\ 
$^{1}$  Centro Brasileiro de Pesquisas F\'\i sicas, \\
Rua Dr. Xavier Sigaud 150 \\
22290-180 Rio de Janeiro -- RJ, Brazil\\
\\
$^{2}$  Universidade Cat\'olica de Petr\'opolis, \\
Bar\~ao de Amazonas \\
25685-070, Petr\'opolis -- RJ, Brazil\\
\\
$^{3}$ Astronomy Unit, School of Mathematical Sciences, \\
Queen Mary, University of London, \\
Mile End Road, London E1 4NS, UK
}

\begin{document}

\date{\today}

\maketitle

\begin{abstract} 
\noindent 
Using recent observational constraints on cosmological 
density parameters, together with recent mathematical results 
concerning small volume hyperbolic manifolds, we argue that, by 
employing pattern repetitions, the topology of nearly f\/lat 
small hyperbolic universes can be observationally undetectable. 
This is important in view of the facts that quantum cosmology
may favour hyperbolic universes with small volumes,
and from the expectation, coming from inf\/lationary scenarios,
that $\Omega_0$ is likely to be very close to one.
\end{abstract}

\vspace{0.8cm}

It is well known that general relativity is a local metrical 
theory and therefore the corresponding Einstein f\/ield
equations do not f\/ix the global topology of spacetime.
This freedom has resulted in a great deal of interest in 
the possibility that the universe may possess compact
spatial sections with a non-trivial topology, which
do not necessarily have positive curvature (see for
example~\cite{CosmicTop}~--~\cite{ZelNov83} and references 
therein). 

Interest in such spaces has also come from quantum cosmology, 
where the existence of the wave function of the universe may 
require f\/initeness of the spatial sections (see for example%
~\cite{SmallVol1}).
Also, in the `tunnelling from nothing' scenario, manifolds 
with small compact spatial sections may be more likely to 
emerge~\cite{Gibbons98}. 
Furthermore, there is a precise sense in which most compact 
$3$-manifolds are hyperbolic~\cite{Thurston82}. 
Together these facts have motivated the study of hyperbolic 
manifolds with non-trivial topology as possible models for 
our universe. 

There has also been two other important developments recently:
one observational, the other mathematical. Regarding the former, 
recent observations seem to indicate that the ratio of the total 
density to the critical density of the universe, $\Omega_0$, is 
likely to be very close to one~\cite{BooMax},
with a signif\/icant proportion of this energy being in the form of 
a dark component with negative pressure~\cite{DarkE}. 

Regarding the latter, we f\/irst of all discuss brief\/ly some 
preliminaries and recall that even though at present there 
is no complete classif\/ication of hyperbolic manifolds, a 
number of important results are known about them, including 
the two important theorems of Mostow~\cite{Mostow} and 
Thurston~\cite{Thurston82}.
According to the former, geometrical quantities of orientable
hyperbolic manifolds, such as their (f\/inite) volumes and the
lengths of their closed geodesics, are topological invariants.
According to the latter, there is a countable inf\/inity of 
sequences of compact orientable hyperbolic manifolds, with the 
manifolds of each sequence being ordered in terms of their
volumes~\cite{Thurston82}, with an overall lower bound which 
is shown to be greater than 0.28151~\cite{Przeworski}.

A natural way to characterize the shape of such manifolds
is in terms of the sizes of their closed geodesics. A particularly 
useful indicator in this regard is the so called {\em injectivity 
radius\/}, $r_{inj}$, def\/ined as the radius of the smallest sphere 
inscribable in $M$.
This in turn allows the def\/inition of a related indicator that has 
often been utilized in most studies regarding searches for topological 
multiple images (see, for example~\cite{SokShv}~--~\cite{grt2001}) 
and references therein), namely the ratio of the injectivity radius
to the depth $\chi_{obs}$ of the astronomical survey up to 
a given redshift $z_{max}$ 
\be
\label{r_inj}
T_{inj}=\frac{r_{inj}}{\chi_{obs}}\;.
\ee
The crucial point regarding this indicator is that, in any universe 
for which $T_{inj}>1$, there would be no observed multiple images 
of either cosmic objects or spots of cosmic microwave background 
radiation (CMBR), and therefore the topology would not be detectable 
observationally using pattern repetition, no matter how accurate 
the observations. Despite the global inhomogeneity of hyperbolic
manifolds this result is location independent 
and therefore applicable to any observer in the universe. 
Similarly the set of universes for which $T_{inj}<1$ are 
observationally detectable through pattern repetitions, at 
least in principle, for some observers.

An important point regarding $r_{inj}$ is that its lower bound 
in the set of all compact hyperbolic manifolds is zero. Thus, 
no matter what the cosmological parameters (and the resulting 
$\chi_{obs}$), there will always exist compact hyperbolic universes  
(with small enough $r_{inj}$ such that $T_{inj}<1$) such that their 
topologies are detectable observationally, at least by some 
observers, even though their number will decrease 
drastically as $\Omega_0 \to 1$~\cite{grt2001}. 

Now given the potential importance of the small volume hyperbolic 
manifolds in connection, e.g.,  with quantum cosmology, the question 
arises as to whether the existence of detectable topologies for all 
values of $\Omega_0 < 1$ still holds if we restrict ourselves only to a 
set of small volume manifolds. 
Surprisingly, the answer turns out to be in the negative. 
To see this, we recall an important set of recent mathematical 
results which show that very small values of $r_{inj}$ do not 
occur in small volume hyperbolic manifolds, and that there is 
in fact a lower bound on the lengths of geodesics in 
any set of small volume closed orientable hyperbolic 3-manifolds%
~\cite{Gabaietal,MarshallMartin}. Thus, for example, according to 
a theorem of Przeworski~\cite{Przeworski}, the shortest geodesic 
in closed orientable hyperbolic 3-manifolds with volume less 
than 0.94274 must have length greater than 0.09, corresponding to 
a lower bound on $r_{inj}$ of $0.045$. An important point for our
purposes here is that it can be shown that there are non-zero 
lower bounds to lengths of shortest geodesics in any set of 
closed orientable hyperbolic 3-manifolds, whose volumes are 
smaller than that of the f\/irst orientable cusped 
manifold by any $\epsilon \neq 0$~\cite{Przeworski2}.

To study an important consequence of these results for cosmology,
let us assume that the universe can be modelled by a $4$-manifold 
$\mathcal{M}$, with a locally hyperbolic isotropic and homogeneous 
Friedmann-Lema\^{\i}tre-Robertson-Walker metric in the standard form 
\be
\label{FLRW1}
ds^2 = -c^2dt^2 + R^2 (t) \left[\,d \chi^2 + \sinh^2 \chi \,(d\theta^2 +
\sin^2 \theta  d\phi^2) \,\right] \;.
\ee
Furthermore,  let the $3$-space be a multiply connected compact
quotient manifold of the form $H^3 /\Gamma$, where $\Gamma$ is a 
discrete group of isometries of $H^3$ acting freely on $H^3$. 
Now in order to have $T_{inj}$, we consider a survey of depth 
up to the redshift $z_{max}$. In this cosmological
setting the depth of the survey expressed in units of the 
curvature radius ($\chi_{obs} \equiv d_{obs}\,/R_0$) 
is given by%
\footnote{For non-flat models the scale factor $R(t)$ is identified with 
the curvature radius of the spatial section of the universe at 
time $t$, and thus $\chi$ can be interpreted as the distance of 
any point with coordinates $(\chi, \theta, \phi)$ to the origin 
of coordinates (in $H^3$), in units of curvature radius, which 
is a natural unit of length and suitable for measuring areas 
and volumes. Throughout this letter we shall use this natural 
unit.}
\begin{equation} \label{X_obs}
\chi_{obs}  =\sqrt{1-\Omega_0}
   \int_0^{z_{max}} \left[ (1+x)^3 \Omega_{m0} +
   \Omega_{\Lambda 0} - (1+x)^2 (\Omega_0 -1) \right]^{-1/2} dx \; ,
\end{equation}
where the current content of the universe is taken to be dust 
(of density $\rho_m$) plus a cosmological constant $\Lambda\,$, 
with $\Omega_m = \frac{8 \pi G \rho_m}{3 H^2}\,$, $\Omega_{\Lambda}=
\frac{\Lambda}{3 H^2}\,$, $\,\Omega = \Omega_m + \Omega_{\Lambda}\,$, 
and where the index $0$ denotes evaluation at present time.

Now to examine the consequences of the above theorem in the light of 
recent cosmological observations, recall that as $\Omega_0 \to 1$, 
the curvature radius increases, resulting in a decrease in 
$\chi_{obs}$, and as a result the set of topologies that would 
be observationally detectable would have to possess $r_{inj}$ 
approaching zero in order to ensure $T_{inj} < 1$. The above 
mathematical results, with their lower bounds on $r_{inj}$, 
have the consequence that the topology of small hyperbolic 
universes would be undetectable for values of $\Omega_0$ close 
enough  to one. The important point being that the range of 
$\Omega_0$ for which the topology of such universes are 
undetectable turns out to be within the range of values 
of $\Omega_0$ allowed by recent observations, and particularly 
those suggested by the inf\/lationary scenarios.

To quantify this, we proceed in the following way. The above 
theorem gives the lower bound on $r_{inj}$ in a set of small 
volume manifolds (those manifolds with volume less than 0.94274, 
which includes at least the Weeks manifold) to be 0.045. This allows 
bounds to be imposed on the ranges of cosmological parameters for 
which the topology is undetectable. We note that 0.045 
is the best estimate available at present and the true lower bound on
$r_{inj}$ may be greater. A way of dealing with this 
possibility is to consider the f\/irst 51 smallest
manifolds of the Hodgson-Weeks census of closed hyperbolic manifolds.
This set contains all manifolds of the census with volumes smaller than
the volume of the f\/irst cusped manifold. We note that the volumes of 
the last 7 manifolds in this set dif\/fer from the volume of the f\/irst 
cusped hyperbolic orientable manifold by a factor which is smaller 
than $10^{-16}$, thus giving an idea of the size of the above 
mentioned $\epsilon$ for the set of manifolds considered here.
The manifold in this set with the lowest $r_{inj} (= 0.152)$ is the 
eighteenth manifold in the census, denoted by $m003(-5,4)\,$,
with volume 1.58865.

Figure~1 gives a plot of the solution curve of equation 
$\chi_{obs} = r_{inj}$ in the $\Omega_{0}\,$--$\,\,\Omega_{\Lambda 0}$
plane for $r_{inj}= 0.045$ and $0.152$, where a survey of depth 
$z_{max}=1200$ (corresponding to the redshift of the surface of 
last scattering, CMBR) was used.
Also included in this Figure is a dashed rectangular box, representing
the relevant part (for our purposes here) of the hyperbolic region of the 
parameter space ($\Omega_0 \in (0.99, 1]$ and $\Omega_{\Lambda 0} 
\in [0.63,0.73]$) given by recent observations~\cite{Bond-et-al-00a}.
For each value of $r_{inj}$ undetectablilty is ensured for the 
values of cosmological parameters (region in the $\Omega_{0}\,$--$\,
\,\Omega_{\Lambda 0}$ plane) which lie above the corresponding curve. 
Thus for $r_{inj} = 0.045$, all closed orientable hyperbolic manifolds 
(universes) with volumes less than 0.94274 would have undetectable 
topology, if the total density $\Omega_0$ turned out 
to be higher than $\sim 0.9998$. 
On the other hand, for $r_{inj} = 0.152$, the topology of none 
of the 51 manifolds of the census (whose volumes range from 0.94271 
to 2.02988 (corresponding to the manifold $m010(4,1)\,$),
and which includes the Weeks manifold) 
would be detectable, if the total density $\Omega_0$ turned out to be
higher than $\sim 0.998$. 
Clearly, the actual precise bounds on $\Omega_0$ depend on 
the precise value of the $\Omega_{\Lambda 0}$ employed.
The resulting allowed changes in the bounds on $\Omega_0$ due to
employing other values in the observed range of $\Omega_{\Lambda 0}$ 
are, however, small, as can be seen from Figure~1.
Similarly, one can easily f\/ind the corresponding ranges of $\Omega_0$ 
for any other particular manifold or f\/inite set of small manifolds. The 
important point is the existence of a lower bound on $r_{inj}$ for 
any f\/inite set of small volume manifolds (smaller, by any 
$\epsilon \neq 0$, than the volume of the f\/irst cusped orientable 
hyperbolic manifold), which in turn 
gives lower bounds on $\Omega_0 <1$, such that the topology of the 
universe is not detectable with methods based on the search for 
pattern repetitions.%
\footnote{Note in addition that, following an assertion by
Thurston which states that the expectation value for the 
length of the smallest closed geodesic at an arbitrary point 
in a generic hyperbolic 3-manifold lies in the range
$0.5 \to 1$, Cornish {\em et al.\/}~\cite{CornSperStar98} 
have argued that even for these large volume hyperbolic manifolds 
$M$, it is very unlikely that the earth is in a region whose closed 
geodesics are as short as the shortest closed geodesics of $M$. 
Thus, according to~\cite{grt2001} the chances of detecting the
topology of these (large) hyperbolic universes are also very low 
according to recent observations ($\Omega_0 \sim 1$), even if 
CMBR is used.} 

To conclude, if it turns out, as suggested by inf\/lationary scenarios, 
that $\Omega_0$ is very close to one, then our results are signif\/icant in 
implying that the topology of small hyperbolic universes, suggested, for
example, by 
some arguments based on quantum cosmology, can be undetectable using 
pattern repetitions. This motivates the development of new strategies 
for looking for the topology of the universe, not based on the 
observation of repeated patterns.

\section*{Acknowledgments}

We are grateful to Andrew Przeworski for very helpful correspondence 
concerning his work, Jef\/f Weeks for many useful comments, and
Neil Cornish for drawing our attention to the reference%
~\cite{CornSperStar98}. We also thank FAPERJ and CNPq for the 
grants under which this work was carried out.

\section*{Figure caption}

\begin{description}
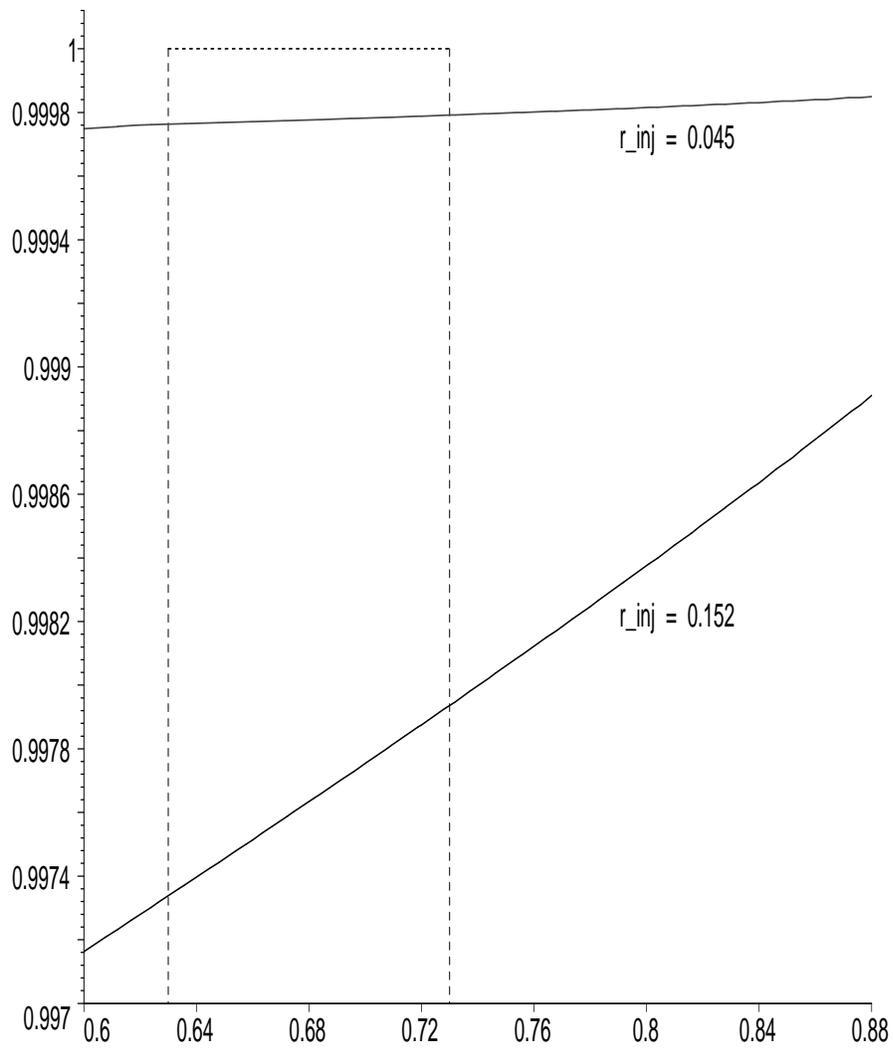
\item[Figure~1]
The solutions curve of $\chi_{obs} = r_{inj}$, as plots of $\Omega_0$ 
(vertical axis) versus $\Omega_{\Lambda 0}$ (horizontal axis),
with $r_{inj}$ taken as $0.045$ (upper curve) and $0.152$ 
(lower curve), respectively. The depth of the survey in both 
cases correspond to a redshift $z_{max}= 1200$ (CMBR). 
Included also is a dashed rectangular box, representing the 
relevant part, for our purposes, of the hyperbolic region of 
the parameter space 
given by recent observations. The undetectable region of the 
parameter space ($\Omega_0, \Omega_{\Lambda 0}$) corresponding 
to each value of $ r_{inj}$ lies above the related curve.
\end{description}


%
\end{document}